\documentclass[aip,graphicx,reprint,floatfix,apl]{revtex4-1}

\usepackage{float}
\usepackage{graphicx}
\usepackage{dcolumn}
\usepackage{chemformula}
\usepackage{gensymb} 
\usepackage{amssymb} 
\usepackage{upgreek}
\usepackage{siunitx}
\usepackage{todonotes}
\usepackage{eqtcommands}
\usepackage{lipsum, babel}

\def\includeSI{1}

\newcommand{\west}{w_\mathrm{est}}
\newcommand{\SuppMat}{Supplementary Material}
\begin{document}


\title{Efficient mapping and tracking the properties of micromechanical resonators using phase-lock loops with closely-spaced frequencies}



\author{Agnes Zinth}
\affiliation{Department of Physics, TUM School of Natural Sciences, Technical University of Munich, Garching, Germany}
\affiliation{Department of Electrical Engineering, TUM School of Computation, Information and Technology, Technical University of Munich, Garching, Germany}
\affiliation{Munich Center for Quantum Science and Technology (MCQST), Munich, Germany}

\author{Samer Houri}
\affiliation{imec, Leuven, Belgium}

\author{Menno Poot}
\email{menno.poot@tum.de}
\affiliation{Department of Physics, TUM School of Natural Sciences, Technical University of Munich, Garching, Germany}
\affiliation{Munich Center for Quantum Science and Technology (MCQST), Munich, Germany}
\affiliation{Institute for Advanced Study, Technical University of Munich, Garching, Germany}

\date{08-12-2025}

\begin{abstract}
Studying the dynamical behavior of micro- and nano-mechanical systems (MEMS and NEMS) is essential in various fields from nonlinear dynamics to quantum technologies. Hence, it is important to be able to precisely monitor the mechanical properties of MEMS and NEMS devices.
In this work, we show how to track and spatially map various properties of a mechanical resonator, such as frequency shift, linewidth, and nonlinearity, by aptly choosing three closely-spaced drive frequencies and using phase-locked loops.
This technique tracks changes in the system faster and more efficiently, without the need for repeated frequency sweeps of the oscillator response, simply by employing three phase-locked tones.
\end{abstract}

\maketitle 
Micromechanical systems have become a vital component of various technologies, including sensing.
This ranges from gas sensors to quantum applications \cite{sanz2022high, sader2018mass, bhattacharya2025two, li2024miniature, bleszynski-jayich_science_persistent_currents}, as well as communication systems \cite{bozkurt2025mechanical},  biomedical instrumentation \cite{arlett2011comparative}, even to applications such as cancer diagnosis \cite{kosaka2014detection}. The performance and reliability of these devices are strongly influenced by their dynamic characteristics, particularly the frequency response, which determines key parameters such as resonance frequency, damping rate, and nonlinearities. Precise frequency response characterization is therefore essential for understanding the system dynamics, monitoring them, and ensuring stable operation under varying conditions. 
Traditional approaches, like recording linewidth maps of 2D material resonators \cite{jaeger2023mechanical}, monitoring magnetic and electronic phase transitions \cite{vsivskins2020magnetic} or optimizing a fabrication process, e.g. the influence of annealing on the quality factor \cite{liu2022effect, aubin2003laser, guneroglu2025study} to characterize mechanical resonators typically involve sweeping the frequency of the driving force over the resonance to measure the amplitude and phase response. While effective, frequency sweeps are time-consuming and prone to drift over time, which limits their ability to track rapid changes. 
To overcome these limitations, frequency tracking can be achieved through phase-locked loops (PLLs). This has, e.g., enabled continuous monitoring of resonance frequency with high precision \cite{blaikie2019fast, Garcia-Sanchez_PRL_Casimir, bleszynski-jayich_science_persistent_currents, yang_NL_masssensing}. Similar to the multi-frequency AFM technique, where either the first two modes \cite{kort2022utilization, santos2023quantification} or higher harmonics \cite{chandrashekar2022sensitivity} are used, we also use several frequencies to gain information about our resonator. 

Here, we present a method of efficiently mapping and simultaneously tracking several properties --- specifically linewidth, frequency, as well as nonlinearity --- of a mechanical resonator. 
We aptly select three closely-spaced frequencies within the width of a resonance peak, follow these with PLLs, and determine the desired quantities from them. 
First, we cover the concept behind our technique and present the 
experimental realization in a micromechanical membrane. We start by tracking the resonators' properties locally during a temperature sweep, then extend to spatially mapping them. Lastly, the nonlinear regime will be explored.

\begin{figure}[htbp]
  \includegraphics[width=1.0\columnwidth]{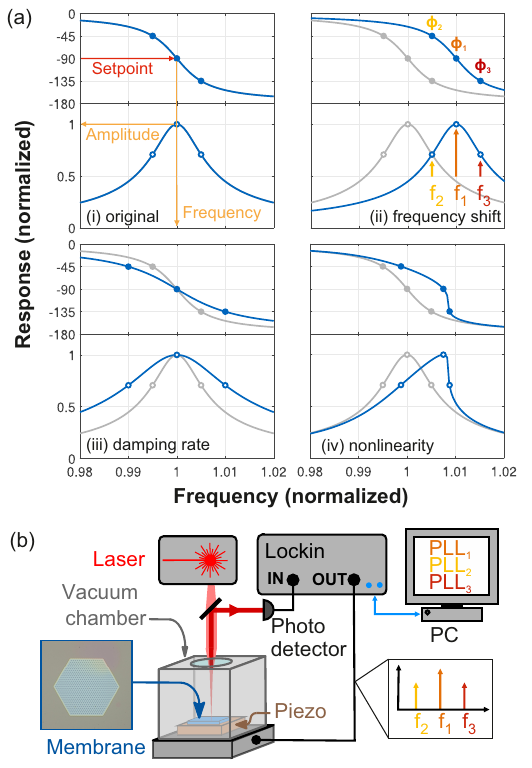}
  \caption{
  (a) Illustration of the concept showing the frequency response of a driven harmonic oscillator with phase (top, in deg.) and magnitude (bottom) for every panel. 
  (i) PLLs can be used to lock to specific phases (filled symbols), resulting in different driving frequencies and amplitudes (open symbols). Now, (ii) a shift in resonance frequency, (iii) damping rate, and (iv) nonlinearity all result in distinguishable shifts in $f_i$. The original response with $w/f_0 = 0.01$ and $\alpha = 0$ is shown in gray. 
  (b) Schematic of the setup with a lock-in amplifier and digital implementation of the PLLs; for details, see Ref.~\onlinecite{hoch_MM_mode_mapping}
\label{fig:intro}}
\end{figure}

For tracking properties of a resonator, phase-locked loops (PLLs) are instrumental \cite{Horowitz,blaikie2019fast,Hippold2024}. A PLL is locked to its setpoint as upon changes in phase, the frequency is adjusted accordingly, as illustrated in Figure \ref{fig:intro}(a) panel (i).
The information can be in the frequency and/or in the amplitude. For example, we used the latter to record mode maps with high resolution, while remaining stable against external influences\cite{hoch_MM_mode_mapping, sommer_ST_xtalk}.
There, we chose one frequency per mode 
to map up to six modes of a membrane simultaneously, but here we use three frequencies within the peak of a \emph{single} mode, see Fig.~\ref{fig:intro}(a). By choosing three setpoints $\phi_{1,2,3}$ --- indicated by the filled symbols --- instead of one per mode, we gain access to properties beyond amplitude and resonance frequency, that are traditionally obtained by frequency scans with a network analyzer (NWA) \cite{denis2018identification}.
With well-chosen setpoints
, the frequency shift, damping rate, and even nonlinearity of the system can be determined quickly, and tracked in real-time and mapped.

For a shift in resonance frequency $f_0$, all three frequencies move in unison as depicted in panel (ii). $f_1$ remains at $f_0$; 
The frequency shift can thus be determined directly from $f_1$. Panel (iii) shows an increase in 
damping.
Now, the center frequency $f_1$ remains unchanged, whereas $f_2$ and $f_3$ move 
apart symmetrically. 
The setpoints $\phi_2 = -45\degree$ and $\phi_3 = -135\degree$ are chosen
, so that $f_3-f_2$ directly yields the second parameter, the linewidth $w$ (see \SuppMat).
Still, with three frequencies, an additional property can be determined. Panel (iv) shows that for increasing nonlinearity --- as quantified by the Duffing parameter $\alpha$ \cite{cleland2013foundations} --- the phase of the response steepens on one side and flattens on the other. Thus, the equality between $f_3-f_1$ 
and $f_2-f_1$ 
no longer persists; this can be used to determine $\alpha$. However, operation in the nonlinear regime is potentially more complex than this simple picture, as we will see later. 
In the following, we first focus on the experimental realization in the linear regime and then return to investigate the nonlinearity.

\begin{figure}[tb]
  \includegraphics[width=1.0\columnwidth]{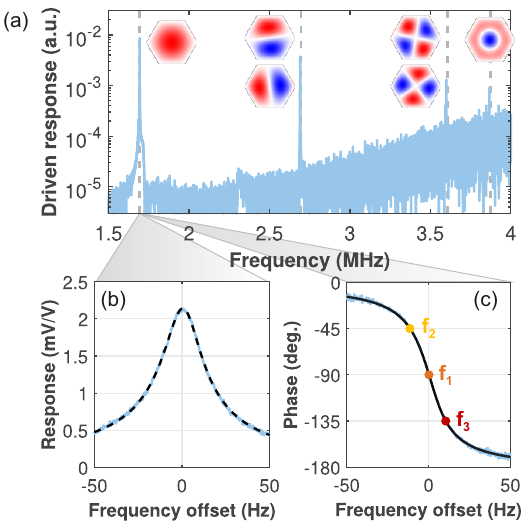}
  \caption{
    (a) Driven response of a hexagonal SiN membrane with resonances near the simulated eigenfrequencies (dashed lines). The corresponding mode shapes are shown as insets.
    (b) Magnitude and (c) phase of the fundamental mode. The setpoints are indicated as symbols.
  \label{fig:membrane}}
\end{figure}

The resonator we use to demonstrate our technique is a hexagonal micromechanical membrane made from high-stress silicon nitride (SiN) \cite{sommer_PRL_nonlinear_map}, as further described in the Methods of the \SuppMat. The sample is mounted on a piezo and excited with a lock-in amplifier as shown 
in Fig.~\ref{fig:intro}(b). This particular membrane has a number of (unintentional) features, like thickness variations or particles on the devices, that are instrumental in showing the applicability of our technique in real-life situations. 
Figure~\ref{fig:membrane}(a) shows the driven response of the membrane in vacuum
. The different drum modes are visible and 
the insets show the simulated mode shapes. 
In the following, we focus on the fundamental mode near $f_0 = 1.69 \un{MHz}$. The zooms [(b),(c)] display an exemplary resonance that is fitted well by a harmonic oscillator response (dashed), yielding a linewidth of $w = 21.79 \un{Hz}$. 
As discussed above, by using three closely-spaced PLLs with setpoints as indicated in (c), the same properties can be extracted without the need to measure and fit the full response.

\begin{figure}[tb]
  \includegraphics[width=1.0\columnwidth]{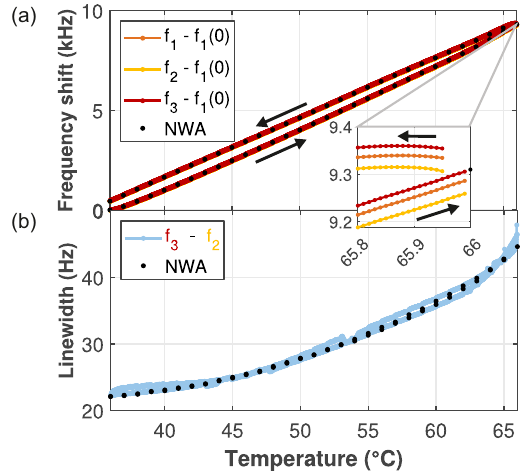}
  \caption{Tracking the (a) Frequency shift and (b) Linewidth during a large temperature ramp. Black data points indicate calibration NWA measurements. The inset in (a) is a zoom in the reverse point of the temperature showing a small offset. }
  \label{fig:temp_sweep}
\end{figure}

The first demonstration of our technique is tracking the resonator properties during a large temperature ramp. The temperature was swept forward and backward from $36$ to $66 \un{\degree C}$ over the course of six hours with the PLLs tracking the changes. 
Figure~\ref{fig:temp_sweep}(a) shows the frequency shift and (b) the linewidth as a function of temperature. NWA calibrations were taken at fixed intervals (black dots) and these match the results from our PLL well; for more details
, see Supp. Sec.~\ref{sec:supp:temp}. Overall, the resonance frequency shifts up by $10 \un{kHz}$: 
The substrate expands more than the membrane, 
resulting in a higher resonance frequency. The frequency shift shows hysteretic behavior as the temperature of the sample mount changes, but the chip temperature lags behind.
This is also the cause for the offset, visible in the inset: The last point of the forward sweep is recorded, an NWA measurement is done $\approx 33\un{s}$ while the setpoint temperature already drops but that of the chip still rises. Afterwards, the first point of the backward measurement starts at a slightly lower temperature but higher frequency. 
\\
The corresponding linewidth in Fig.~\ref{fig:temp_sweep}(b) has more than doubled. This is due to an increased coupling of the membrane to substrate modes  \cite{Joeckel_PRL_spectroscopy_Q_SiN_membrane}. The determined values again nicely match the NWA measurements, but extracting the linewidth precisely from the NWA traces requires some tricks in the data processing, see \ref{sec:supp:temp}. Still, 
the measurement time is far longer than for the PLL technique. 

A closer look at the inset in Fig.~\ref{fig:temp_sweep}(a) reveals something unexpected: there is an asymmetry in the frequency differences. This is because the PLLs needs time to follow rapid changes. The asymmetry of $f_2$ and $f_3$ with respect to $f_1$ arises due to the dependence of the PLL response on the shape of the phase response, see \SuppMat. 
To compensate for this, we use an estimator in the post-processing of the data, taking into account the tracking error of the PLL. The three phases $\phi_i$ and corresponding frequencies $f_i$ form a linear system of equations that can be solved giving more accurate estimations for $f_0$ and $w$; the full derivation and comparison of estimator vs. subtraction can be seen in Supp.~\ref{supp:subsec:est}.

From measuring at one position on the membrane, we extend to spatially mapping its properties. However, as neither systematic frequency shift nor linewidth change with position is expected in our membrane, we introduce these by implementing a feedback loop that feeds the measured displacement back 
with a controllable gain $g_\mathrm{FB}$ and phase shift $\theta_\mathrm{FB}$ \cite{poot_APL_cooling, poot_NJP_Yfeedback} as detailed in \SuppMat. 
A feedback force that is in phase with the displacement shifts the frequency, whereas feedback that is in phase with the velocity alters the damping, i.e. the linewidth $w$ \cite{poot_physrep_quantum_regime}. Since the total loop gain depends on $g_\mathrm{FB}$, as well as on the open loop response (varying on the micrometer scale due to the release holes, and slower due to the mode shape and readout), this results in both fine and large-scale features in the local frequency shift and damping rate.

\begin{figure}[tbh]
  \includegraphics[width=\columnwidth]{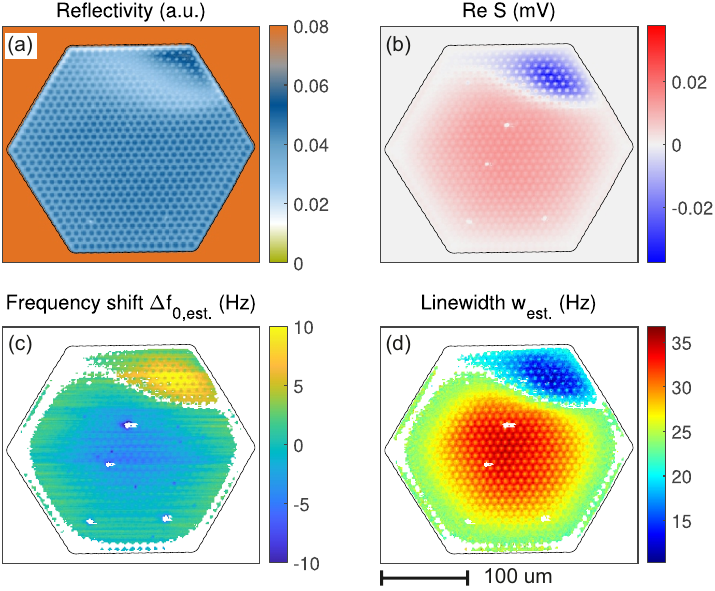}
  \caption{Maps in the presence of feedback ($g_\mathrm{FB} = 250 \un{V/V}$, $\theta_\mathrm{FB} = -45 \un{^o}$). (a) reflected laser light, (b) signal amplitude, i.e. the product of the mode shape and transduction (see Ref.~\onlinecite{hoch_MM_mode_mapping} for details on the meaning of $\mathrm{Re}~S$). (c) the frequency shift and (d) the estimated linewidth. The black line is a contour of the reflectivity and outlines the edge of the membrane. White in (c) and (d) are excluded regions where the signal is too low for the PLL to update the frequencies, or the error for the estimator is too large. 
  \label{fig:map_FB}} 
\end{figure}

Figure~\ref{fig:map_FB} shows the spatial map of various properties in the presence of 
feedback: (a) shows the reflectivity and (b) the amplitude of PLL 1. The latter deviates from the expected mode shape [Fig.~\ref{fig:membrane}(a)], as the membrane has a region with a different SiN thickness where sign of the readout is reversed (Fig.~\ref{fig:supp:map_noFB}) \cite{zinth_membrane_readout}. Figure~\ref{fig:map_FB}(c) shows the frequency shift, (d) the estimated linewidth. White spots in any of the maps are contamination on the membrane where the laser heats locally, shifting $f_0$ down sharply and reducing the signal. These points are either excluded due to low signal or large errors. Yet, the PLL tracking remains  robust against such real-world features, including the sign reversal. In the center of the membrane, the frequency shifts $\approx 10\un{Hz}$ downwards with respect to the edge, in the upper right corner upwards as the loop gain switches sign. Overall, a very fine structure is visible, and even the holes in the membrane can be resolved. A similar pattern occurs in the linewidth (d). There, the linewidth increases in the center of the membrane and decreases in the upper right corner, again due to the feedback. Without it, the linewidth is constant at $\approx 22 \mathrm{Hz}$ 
on the whole membrane (Fig.~\ref{fig:supp:map_noFB}). 
As each point only takes $\lesssim 2 \un{s}$ (mainly limited by stage motion), mapping such fine structures in $w$ and $f_0$ can be done much faster with our three-tone PLL than by recording a whole NWA trace for every single point over the whole membrane \cite{davidovikj_NL_graphene_mode_visualization, hoch_MM_mode_mapping}, which would take weeks to complete.

\begin{figure}[tbh]
  \includegraphics[width=\columnwidth]{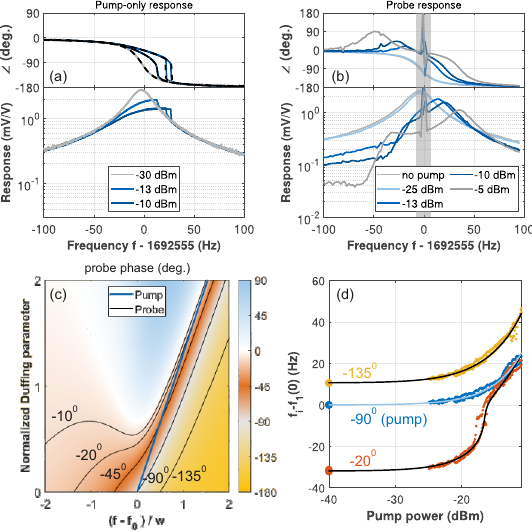}
  \caption{Operation in the nonlinear regime. 
  (a) Driven responses for varying excitation power. The dashed lines show the fit of the phase.
  (b) Response of a weak (-30 dBm) probe for different pump powers. The pump at 1692555 Hz is grayed out.
  (c) Calculated probe response phase for different probe frequencies and strength of the nonlinearity. Contour lines at the indicated values are shown in black; the blue line shows the frequency of pump when locked to $-90\degree$.
  (d) Evolution of the PLL frequencies for increasing pump strength. The pump is in blue and the two probes are gold and orange. For clarity, the markers at -40 dBm have been enlarged. The solid lines are the scaled contours from (c).
  \label{fig:nonlinear}} %
\end{figure}

The final property that can be determined with our technique is the nonlinearity. Thus far, the membrane was driven in the linear regime; increasing the driving gives access to the nonlinear response. In Figure~\ref{fig:nonlinear}(a) the driven response for different driving powers with only \textit{one} tone (``pump'') is shown. When the drive increases, the 
resonance takes the shape of a Duffing response. Although the amplitude is affected by nonlinear readout, the phase --- used by the PLLs --- is not \cite{sommer_PRL_nonlinear_map}. Its slope gets steeper with increasing excitation power, as was shown in Fig.~\ref{fig:intro}(a) panel(iv) so an asymmetry between $f_3-f_1$ and $f_1-f_2$ could be expected. Unfortunately, the nonlinearity makes the situation more complex: to apply our technique, we require three driving tones, but already adding a second, strong tone gives rise to very complex effects \cite{Houri_JJAP_pulse_width_duffing, catalini_arXiv_detuned_tones}. Hence, instead of equally strong tones for all PLLs, we use weak probes \cite{Antoni_EPL_nonlinear_membranes}. Figure \ref{fig:nonlinear}(b) shows the responses of one probe 
for varying pump powers. Starting with one relatively weak pump tone of $-25\un{dBm}$ at a fixed frequency, the probe response is close to the original linear response, only slightly shifted compared to the case of no pump (gray).
Increasing the pump power leads to more complex probe responses. This is still captured by an analytical expression \cite{defoort_APL_selfcoupling}, which also shows that adding another weak probe can be done independently, see \ref{supp:nonlinear}. For frequencies higher than the pump, the measured amplitude and phase in Fig.~\ref{fig:nonlinear}(b) look like a shifted linear response; the stronger the pump, the larger the shift. For lower frequencies, the response becomes more complex, with the phase developing a peak with positive values. Figure~\ref{fig:nonlinear}(c) shows a color plot of the calculated probe response phase   
versus nonlinearity induced by a $-90\degree$-locked strong pump. Note that, PLLs would follow the contour lines. Like the traces in (b), all setpoints shift towards higher frequencies. The pump at $-90 \degree$ is indicated in blue; the $-135\degree$ line remains on the right side of the pump, hence we keep that setpoint for $\phi_3$. 
However, the previous $\phi_2 = -45\degree$ would cross the pump relatively fast with increasing pump power. Hence, PLL 2's lock would already be lost for small nonlinearities, due to the frequencies approaching each other, and a different setpoint has to be chosen. Indicated in Fig.~\ref{fig:nonlinear}(c) are $-10\degree$ and  $-20\degree$. The trend of the former does not behave monotonously anymore and would therefore lead to the loss of the PLL lock for increasing or decreasing pump power. Contrary, the $-20\degree$ setpoint is farther away from the pump than $-45\degree$ and can therefore track higher pump powers, yet remaining monotonous. 
\\
The measurement with the adjusted values is shown in Fig.~\ref{fig:nonlinear}(d). The black lines indicate the scaled lines from (c), while the different colors are measured data points, which match nicely. In the linear regime, for small pump powers, the PLL frequencies evolve in unison, then they start shifting upwards as the response enters the Duffing regime. Above $-20\un{dBm}$, a rather sharp transition is reached and the $-20\degree$ setpoint approaches the pump frequency, but still matching the calculations. 
There is a 
limit in the experiment on how close two frequencies can get, but, importantly, from the scaling of the contour lines, not only $f_0$ and $w = 22 \un{Hz}$ are obtained, but also the amount of nonlinearity. Specifically, we find a critical pump power of $P_c = -12.7 \un{dBm}$, which matches well with the independently-determined value of $-12.57 \un{dBm}$ (\SuppMat). So 
not only the linear properties, but also the nonlinearity of the strongly-driven resonator can be determined with our method. 

In conclusion, we demonstrate a fast and efficient method for tracking and mapping properties of a mechanical resonator, like the frequency shift, linewidth and nonlinearity using three strategically-chosen driving frequencies instead of repeatedly recording frequency sweeps. This very robust method allows to investigate a plethora of unintentional features of our resonator like thickness variations, particles on the membrane or the linewidth dependence on temperature. Thus, it enables a fast characterisation, spatial mapping and monitoring of properties of changing systems in real time. 

\section*{Supplementary Material}
The Supplementary Material contains details on the methods used, the fabrication of the silicon nitride membranes and the setup with the implemented feedback loop. Further included is the theoretical background, supporting data on the on the temperature ramp measurement and reference measurements without applied feedback. Finally the operation in the nonlinear regime is covered more thoroughly. 

\begin{acknowledgments}
We thank Timo Sommer and Johannes Daniel for assistance with nanofabrication. Funded by the German Research Foundation (DFG) under Germany's Excellence Strategy - EXC-2111-390814868 and TUM-IAS, funded by the German Excellence Initiative and the European Union Seventh Framework Programme under grant agreement 291763.\end{acknowledgments}

\section*{References}
\bibliography{membranes}

\ifx\makeDiff\undefined
  \ifx\includeSI\undefined
  \else
    \def\calledSI{1}
\ifx\calledSI\undefined
   
    \documentclass[aip,graphicx,reprint,floatfix]{revtex4-1}

    \usepackage{float}
    \usepackage{graphicx}
    \usepackage{dcolumn}
    \usepackage{chemformula}
    \usepackage{gensymb} 
    \usepackage{amssymb} 
    \usepackage{hyperref}
    \usepackage{upgreek}
    \usepackage{siunitx}
    \usepackage{eqtcommands}
    \newcommand{\west}{w_\mathrm{est}}

    \begin{document}
    \title{Efficient mapping and tracking the properties of  micromechanical resonators using closely-spaced phase-lock loops}
    
    \author{Agnes Zinth}
    \affiliation{Department of Physics, TUM School of Natural Sciences, Technical University of Munich, Garching, Germany}
    \affiliation{Department of Electrical Engineering, TUM School of Computation, Information and Technology, Technical University of Munich, Garching, Germany}
    \affiliation{Munich Center for Quantum Science and Technology (MCQST), Munich, Germany}
    
    \author{Samer Houri}
    \affiliation{imec, Leuven, Belgium}

    \author{Menno Poot}
    \email{menno.poot@tum.de}
    \affiliation{Department of Physics, TUM School of Natural Sciences, Technical University of Munich, Garching, Germany}
    \affiliation{Munich Center for Quantum Science and Technology (MCQST), Munich, Germany}
    \affiliation{Institute for Advanced Study, Technical University of Munich, Garching, Germany}

    \date{\today}
    \begin{abstract}
    \end{abstract}
    
    \maketitle 
    \onecolumngrid
\else
    \onecolumngrid
    \section*{Supplementary Material}
\fi

\setcounter{section}{0}
\renewcommand\thesection{S\arabic{section}}
\renewcommand\thesubsection{\arabic{subsection}}

\setcounter{figure}{0}
\renewcommand\thefigure{S\arabic{figure}}
\setcounter{equation}{0}
\renewcommand\theequation{S\arabic{equation}}
\setcounter{table}{0}
\renewcommand{\thetable}{S\arabic{table}}

\noindent In this Supplementary Material, we show supporting measurements and data, and details about the methods used, including fabrication of the membranes and the measurement setup. Further, we give the theoretical background covering the harmonic oscillator response function, the calculation of the estimated quantities and the PLL's asymmetry in the temperature ramp measurement. Lastly, the operation in the nonlinear regime is covered.
\\ \\

\section{Methods} \label{sec:supp:setup}
\noindent Membranes similar to those in Ref.~\onlinecite{adiga_APL_SiN_drum_Q_mode} were made as described by Hoch \emph{et al.} \cite{hoch_MM_mode_mapping} by lithographically defining and dry-etching release holes in 330-nm-thick high-stress SiN, followed by an isotropic wet etch of the now-exposed silicon oxide using buffered hydrofluoric acid. A \SI{130}{\minute} release etch results in the formation of a fully suspended SiN membrane above a flat Si surface \cite{adiga_APL_SiN_drum_Q_mode, hoch_MM_mode_mapping}. The oxide etch also removes part of the SiN, making the membrane thinner than the original thickness of the SiN thin film  \cite{sommer_APL_membrane_AlN, poot_ST_AlN_simulations}. Unlike our previous studies with rectangular membranes and a square lattice \cite{poot_ST_AlN_simulations}, this membrane has a triangular lattice of release holes and a hexagonal outline \cite{sommer_PRL_nonlinear_map}.

We use our interferometry setup [Fig. 1 (b)] which is detailed in our previous work \cite{hoch_MM_mode_mapping, poot_ST_AlN_simulations, sommer_ST_xtalk} to measure the vibrations of the membrane in vacuum. In summary, a HeNe laser is focused on the membrane and the probing location is set with a stepper-motor-driven xy-translation stage. The sample is mounted on top of a Peltier element to control the temperature, which heats the whole chip. Except for the temperature-ramp measurement, the temperature is kept stable at $36\un{^{\circ}C}$. When the membrane moves, the reflectivity changes and is recorded by the photodetector (interferometric readout). A bias-tee splits the electrical signal from the photo detector into a d.c. part, which is used for the reflectivity measurement, and an a.c. part containing the vibration signal. Vibrations are excited with a piezoelectric element underneath the sample, which is driven by the output of the lockin amplifier (Zurich Instruments HF2). This lockin can also be used to record frequency sweeps, as a NWA would. We use a sweep over a large frequency range to determine the resonance frequency [Fig.~2(a)].  

Experimentally, the measured response is shifted compared to $\angle H$ due to electrical delays and the excitation with the piezo. This device- and setup-dependent shift is taken into account in the actual values set for the PLLs in our experiments. From an initial frequency sweep, we find about $-90\degree$. For clarity, we always give phases with this shift subtracted, i.e. the nominal setpoints of the three PLLs are $-90\degree$, $-45\degree$ and $-135\degree$, respectively. Also the offset is subtracted from the data presented in this work [e.g. in Fig.~2(c)] so that it directly compares to the theoretical phase response. 

As in our previous works on efficient mode mapping \cite{hoch_MM_mode_mapping, sommer_ST_xtalk}, up to 6 PLLs (here, only 3 are used) are implemented digitally in LabVIEW by processing the signals from the lockin and setting its output frequencies (which equal the demodulation frequencies) using a proportional-integral control loop. Cross talk \cite{sommer_ST_xtalk} is not significant in the measurements presented here. Care is taken that the demodulator bandwidth is narrow enough so that the nearby frequencies are not picked up significantly. 
Post processing of the measurement data is done using {\sc Matlab}.

\begin{figure}[tbh]
  \includegraphics[width=.7\textwidth]{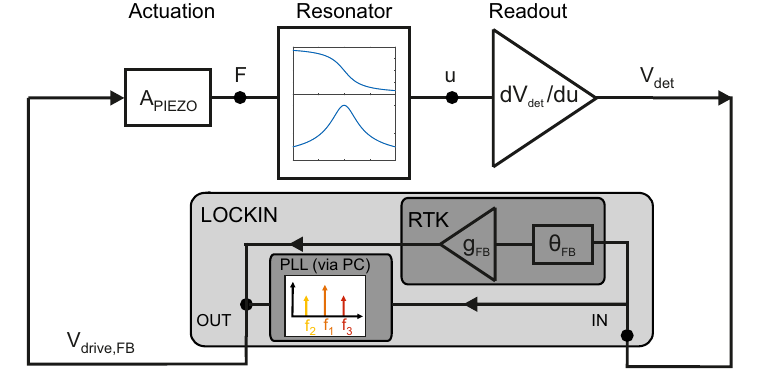}
  \caption{Sketch of the feedback part in the setup. The lockin outputs the drive for the three PLLs and a program is uploaded to the internal processor (RTK) to generate the feedback signal. The output actuates the piezo, resulting in the displacement of the membrane. The change in reflected power is detected by the photodetector.}
  \label{fig:supp:FB_setup}
\end{figure}

\begin{figure}[tbh]
  \includegraphics[width=\textwidth]{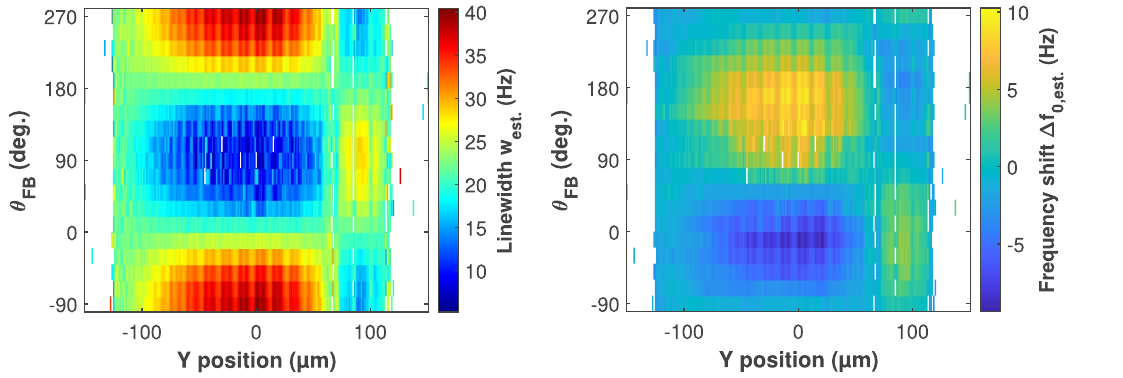}
  \caption{Color maps of the estimated linewidth (left) and frequency shift (right) as a function of the feedback phase $\theta_\mathrm{FB}$ while stepping over the membrane with a feedback gain of $g_\mathrm{FB} = 250 \un{V/V}$. 
  Note how the feedback-induced shifts reverse sign when the transduction becomes negative beyond $y \gtrsim 70 \un{\mu m}$. 
  \label{fig:supp:FB}}
\end{figure}

For the measurements in Fig. 4, a feedback loop is added to the setup, as schematically shown in Fig. \ref{fig:supp:FB_setup}. The feedback is implemented using the digital-signal processor of the lockin \cite{poot_NJP_Yfeedback, poot_PRA_squeezing_feedback}, namely the Real Time Kit (RTK). It updates at 14\,400 Sa/s and has a bandwidth of 5000 Hz around 1\,692\,550 Hz. The lockin outputs the feedback signal (FB) as well as the drive for all three PLLs. This is forwarded to the piezo, actuating the mechanical resonator with an inertial force $F$, and read out via the reflected light. The gain $g_\mathrm{FB}$ is stated in V/V so that the total loop gain is given by the product of the (open-loop) driven response $y(f)$ and the feedback gain including the feedback phase shift $\theta_\mathrm{FB}$.
Like the PLL setpoints, the stated feedback phase is corrected for the overall shift so that $\theta_\mathrm{FB} = -90 \un{^o}$ corresponds to an increased damping rate \cite{poot_APL_cooling} - at least for a positive transduction - as can be seen in Fig.~\ref{fig:supp:FB}. See Ref.~\onlinecite{poot_NJP_Yfeedback} for details on a similar implementation of feedback using the lockin amplifier.

Fig. \ref{fig:supp:FB} shows the estimated linewidth and frequency as a function of the feedback phase $\theta_\mathrm{FB}$ for different positions in the y-direction. The feedback gain is kept constant at $g_\mathrm{FB} = 250 \un{V/V}$. When the transduction becomes negative, the shifts induced by the feedback reverse sign. This is visible from  $y \gtrsim 70 \un{\mu m}$, i.e., in the region where $\frac{\mathrm{dV}}{\mathrm{du}}$ is reversed due to thickness variation of the SiN (see also Fig.~\ref{fig:supp:map_noFB}). 

\section{Temperature ramp measurement} \label{sec:supp:temp}
\begin{figure}[tbh]
  \includegraphics[width=9cm]{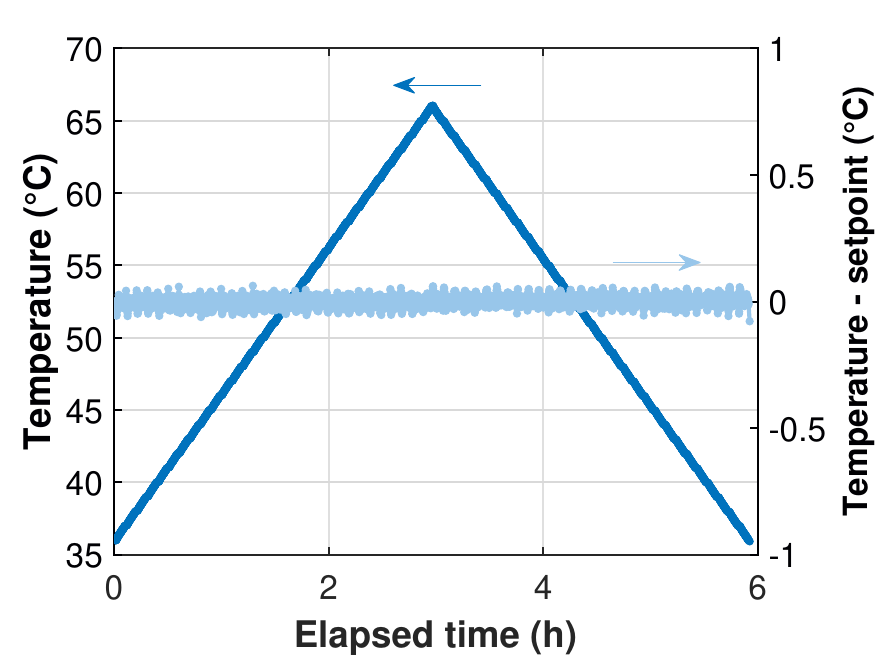}
  \caption{Measured temperature (left) and difference with the setpoint (right) as function of the time since the start of the measurement of Fig. 3.
  \label{fig:supp:T_vs_t}}
\end{figure}

\noindent Figure~\ref{fig:supp:T_vs_t} shows the evolution of the measured temperature of the measurement shown in Fig. 3 of the main text. The thermometer (Pt-1000) is located inside the mounting plate for the PCB with the piezo and chip. This aluminum plate sits on top of a Peltier heater, which is controlled via the PC. The measured temperature follows the setpoint of the temperature control closely, but the sample temperature may be lagging behind due to the thermal resistance of the PCB and piezo on which it is glued, causing the hysteresis visible in Fig.~3(a).

\begin{figure}[tbh]
  \includegraphics[width=\columnwidth]{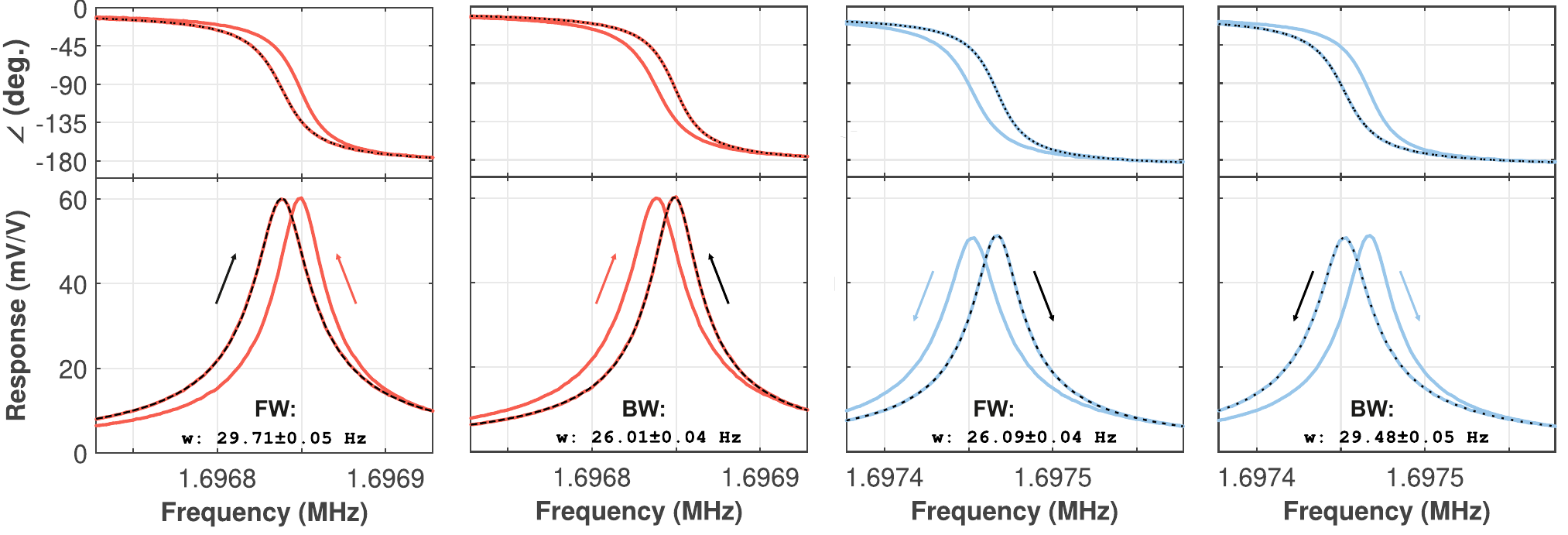}
  \caption{NWA traces at $50\un{^oC}$ for increasing temperature (a),(b) [red] and while cooling (c),(d) [light blue] with fits [black]. The data in (a),(b) and in (c),(d) is identical, but either the forward (a),(c) or the backward (b),(d) direction of the NWA trace is fitted with a harmonic oscillator response (black). The resulting linewidth $w$ is indicated. All NWA traces have a span of 200 Hz and 101 points per sweep direction (202 in total).
  \label{fig:supp:NWA_T}}
\end{figure}

The temperature ramp of $\sim \pm 10 \un{K/h}$, in combination with a temperature coefficient $\sim +3  \un{kHz/K}$ and a linewidth $w \sim 22 \un{Hz}$, means that during the $\sim 33 \un{s}$ that it takes to measure a network trace, the resonance shifts significantly, as shown in Fig.~\ref{fig:supp:NWA_T}. Also, the apparent resonance widens or narrows, depending on whether the sweep direction and frequency shift direction coincide, or not. Without considering this, the linewidth extracted from NWA measurements would thus not be correct. Hence, for the NWA data in Fig. 3, the average value of the fit parameters in the forward and backward driven response frequency sweep are used. Moreover, the large frequency shift of almost 10 kHz means that either the span has to at least as large, which --- in combination with the narrow linewidth --- implies a very long measurement, or the center frequency has to be well chosen. For every NWA calibration (done every 100 measurements, roughly every 6 min.), the latest value of $f_1$ was used for the center of the NWA span. In other words, the PLL have to support the NWA measurements.

Finally, we remark that the temperature sweep was measured after an improvement to the setup, which resulted in larger absolute signals and a different phase offset compared to the other measurements presented in this work.

\section{Map without feedback}
\begin{figure}[tbh]
  \includegraphics[width=\columnwidth]{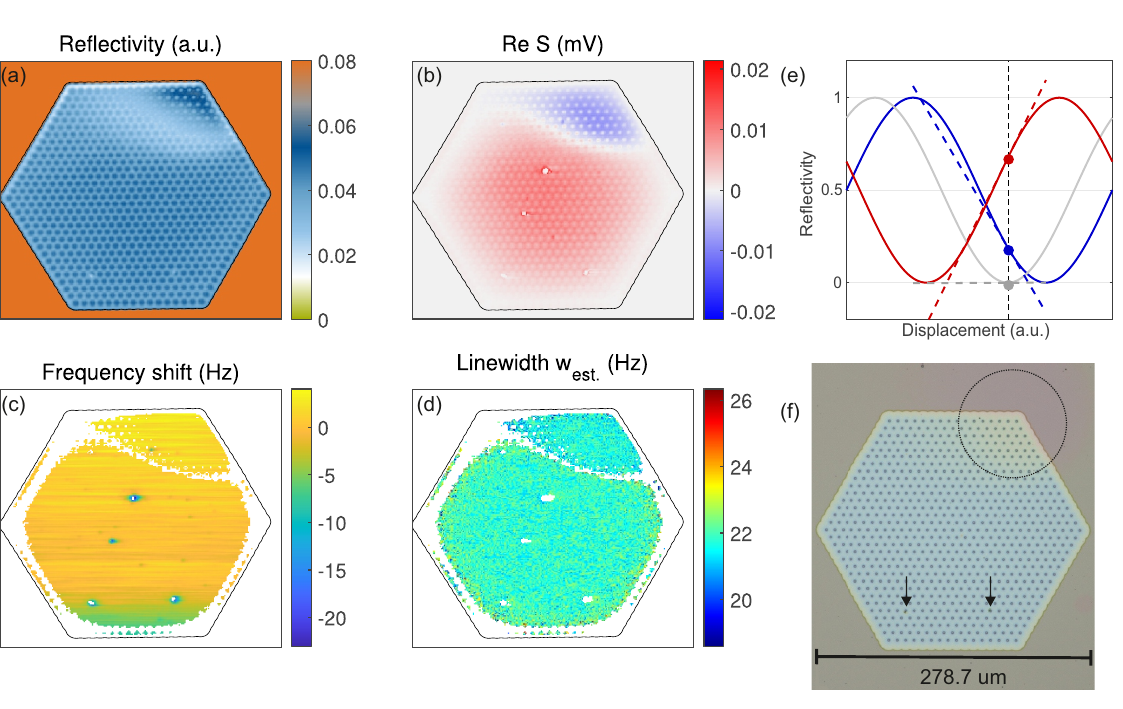}
  \caption{Maps without feedback a) reflectivity, b) amplitude, c) the frequency shift, d) estimated linewidth, e) schematic reflectivity vs. displacement for different SiN thicknesses to explain the sign change in the transduction, f) optical microscope picture indicating the dimension of the membrane and the spot in the upper right corner. The arrows indicate particles on the membrane.\label{fig:supp:map_noFB}} 
\end{figure}
\noindent Figure~\ref{fig:supp:map_noFB} shows mapping measurement as in Fig. 4 of the main text, but now without feedback.
The slightly thicker SiN\footnote{Our working hypothesis is that this is because at that location the SiN is protected against the oxide etchant for some time due to an air bubble or organic residue.} on the upper right corner gives a different sign of the transduction \cite{hoch_thesis, zinth_membrane_readout}, and is also visible as discolorations in the micrograph shown in panel Fig.~\ref{fig:supp:map_noFB} (e) --- both in the supported (brown $\rightarrow$ purple) and in the suspended (light blue $\rightarrow$ yellow) parts. The light gray area separating the blue and red regions in the mode map (b) coincides with the minimum in the reflectivity map (a). This is consistent with a SiN-thickness-change-induced shift of reflectivity-versus-displacement-fringe $R(u)$ from the side with positive $dR/du$, through a minimum, to the side with negative transduction \cite{sommer_PRL_nonlinear_map, zinth_membrane_readout} as illustrated in Fig.~\ref{fig:supp:map_noFB}(e). 
The detected signal (Fig. \ref{fig:supp:FB_setup}) can be described as: 
\begin{equation}
    V_{\mathrm{det}} = \frac{\partial V_{\mathrm{det}}}{\partial P_{\mathrm{det}}} \, P_L \, R(u),
\end{equation}
with $P_{\mathrm{det}}$ the power of the reflected laser light impinging on the photodetector, $P_L$, the laser power, and $R(u)$ the displacement-dependent reflectivity. The transduction is then given by: 
\begin{equation}
    \frac{\partial V_\mathrm{det}}{\partial u} = \frac{\partial R(u)}{\partial u} \frac{\partial V_\mathrm{det}}{\partial P_\mathrm{det}}  \, P_L .
\end{equation}
As a function of displacement, everything included in the readout, $\frac {\partial V_\mathrm{det}}{\partial P_\mathrm{det}}$ --- as well as the laser power $P_L$ --- remains constant. Therefore, only the changes in the transduction are dependent on the change in reflectivity with respect to the displacement. The schematic in Fig. \ref{fig:supp:map_noFB}(e) shows this effect. Every sinusoidal curve $R(u)$ depicts a different SiN thickness. The three dots indicate three different thickness regions at the same displacement. At each point, the reflectivity signal has a different slope (dotted lines): either positive (red), zero (gray) or negative (blue). Each location would have a different sign in the transduction which modulates the mode map in (b). This explains the sign reversal, despite the fundamental mode not having a nodal line \cite{zinth_membrane_readout}. Demonstrating the operation of our multi-tone PLL method in the presence of this and other features shows the robustness in real world applications.
\\
Without the feedback, the (estimated) linewidth map in (d) is constant over the membrane. The frequency $f_0(x,y) - \langle f_0 \rangle$ shows a slow drift of a few Hz (green $\rightarrow$ yellow) as well as a number of localized spots where the frequency drops sharply. These coincide with regions of low reflectivity $R$ and signal $S$, supporting that these are dirt particles/debris that absorb part of the laser light, thereby heating the membrane locally, lowering its frequency \cite{Shaniv_PRR_brownian, Joeckel_PRL_spectroscopy_Q_SiN_membrane}. Near those locations, $w$ remains constant and there is no estimate available exactly at the particle position (white). The particles are not very well visible in the micrograph; only the two spots at the bottom can (barely) be seen and are indicated by the black arrows. 

\section{Theoretical Background}
\label{supp:sec:theory}
In this section, we cover the theoretical background of the harmonic oscillator response function, how the estimated linewidth is determined, and how asymmetry in the PLL tracking arises.

\subsection{Harmonic oscillator response function} \label{supp:subsec:HO}
\noindent When driving our resonator --- the fundamental out-of-plane mode of the hexagonal membrane --- at low powers, it behaves as a linear harmonic oscillator. The amplitude is given by the magnitude $|H(f)|$ of the frequency response function $H(f)$, whereas its phase $\angle H(f)$ quantifies the delay between the drive and resulting motion. For simplicity, we focus on the case where the linewidth $w$ is much smaller than the natural frequency $f_0$ and $|f-f_0| \ll f_0 $ so that we can apply the following approximation
\begin{equation}
    H(f) = \frac{1}{(2\pi)^2\,m} \frac{1}{f_0^2- f^2 + iw f}     \approx \frac{1}{(2\pi)^2\,m} \frac{1}{-2f_0(f-f_0) + iw \,f_0} = \frac{1}{2 (2\pi)^2\,m\, f_0 } \frac{f_0-f-\frac{iw}{2}}{(f_0-f)^2+(\frac{w}{2})^2}.
    \label{eq:supp:L}
\end{equation}
Note that in this Lorentzian approximation, $w$ equals the full-width-at-half-maximum of $|H(f)|^2$. Utilising this fact, we will select our second $-45\degree$ and third $-135\degree$ PLL setpoints exactly at the FWHM of the response function, in addition to the usual $-90\degree$ setpoint. Further, $w = \gamma/2\pi$, where $\gamma$ is the damping rate. These properties are visualised in Fig. 1 (a). 
The phase of the Lorentzian approximation $\angle H(f)$ is given by 
\begin{equation}
     \phi = \angle H(f) = - \arctan \Biggl( \frac{\frac{w}{2}}{f-f_0} \Biggl).
\end{equation}
As the phase $\phi$ is a measured quantity in our experiment, we want to utilize it to determine the linewidth and frequency from it. Rearranging the equation to calculate the linewidth $w$ gives
\begin{equation}
    w = 2\,(f-f_0)\,\tan(-\phi) 
\end{equation}
or
\begin{equation}
    f = f_0 + \half w \cot(\phi). \label{eq:supp:linear}
\end{equation}
Choosing three setpoints also gives access to three phases, which we can use to determine our resonators' properties. Specifically, for $\phi_1 = -90\degree$, $f_1 = f_0$. For $\phi_2 = -45\degree$, $f_2 = f_0 - w/2$ and for $\phi_3 = -135\degree$, $f_3 = f_0 + w/2$ so that $f_0 = f_1$ and $w = f_3-f_2$.
The implementation of a more advanced estimator is covered in the following section. 

\subsection{Estimator} 
\label{supp:subsec:est}
\begin{figure}[tbh]
  \includegraphics[width=1\columnwidth]{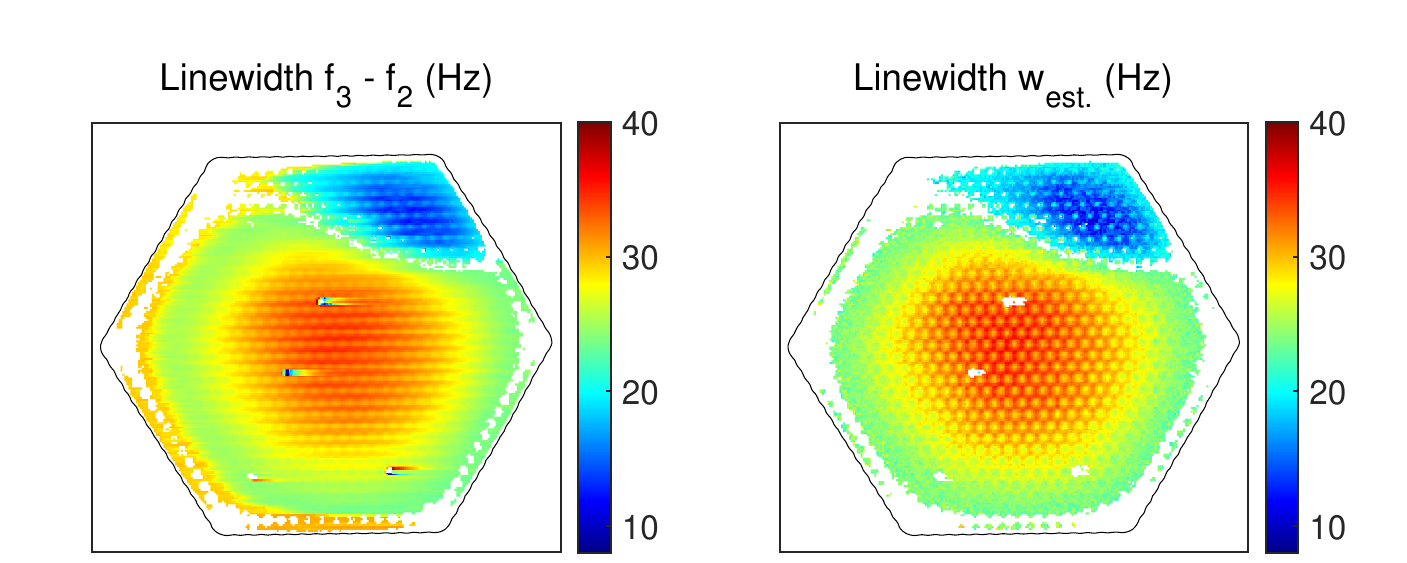}
  \caption{Comparison of the linewidth extraction from the same data using two methods. left: $f_3 - f_2$, right: estimator $\west$. The measurement is done by stepping left to right and then upward, line by line.
  Feedback settings: $g_\mathrm{FB} = 250 \un{V/V}$ and $\theta_\mathrm{FB} = -90 \un{^o}$. 
  \label{fig:supp:west}} 
\end{figure}
\noindent As introduced in the previous section, Sec. \ref{supp:subsec:HO}, we know the phase for each of the three setpoints $\phi_{1,2,3}$. Up to now, there was no distinction between the setpoint of the PLL and the actual phase. From the experiment, we know the measured phase consists of the PLL setpoint $\phi_{i,setpoint}$ and its error $e_i$
\begin{equation}
    \phi_i = \phi_{i,setpoint} + e_i
\end{equation}
With these three phases $\phi_i$, we can set up a linear equation system based on Eq.~\eqref{eq:supp:linear}, allowing us to determine the estimated linewidth ${w_\mathrm{est}}$ and frequency $f_{\mathrm{est}}$. This system of three linear equations and two unknowns is solved (as the least squares solution) by {\sc Matlab}'s \textbackslash-operator: 
\begin{equation}
    \begin{bmatrix}
        \frac{1}{2}w_{\mathrm{est}} \\ f_{\mathrm{est}}
    \end{bmatrix}
     = 
    \begin{bmatrix}
        \cot \phi_1 & 1 \\  \cot \phi_2 & 1 \\  \cot \phi_3 & 1 \\
    \end{bmatrix}
    \Big \backslash
    \begin{bmatrix}
         f_1 \\ f_2 \\ f_3
    \end{bmatrix}
\end{equation}
The result contains the estimated frequency $f_{\mathrm{est}}$ and linewidth $w_{\mathrm{est}}$. As two parameter would be enough to determine frequency and linewidth, the third can be used to determine an additional property or simply for a consistency check. 

\begin{figure}[tbh]
  \includegraphics[width=1.0\columnwidth]{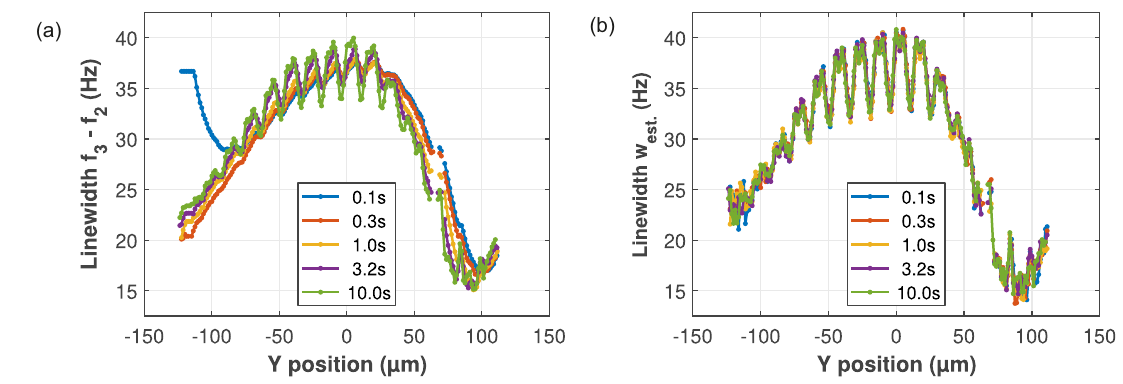}
  \caption{Comparison of the linewidth extraction methods for different wait times between steps: (a) $f_3-f_2$ and (b) the estimator $\west$. Feedback settings: $g_\mathrm{FB} = 250 \un{V/V}$ and $\theta_\mathrm{FB} = -90 \un{^o}$.
  \label{fig:supp:wait}} 
\end{figure}

\noindent Figure~\ref{fig:supp:west} shows the estimated linewidth $w_\mathrm{est}$, and compares the maps of the linewidth in the presence of feedback using two methods. Note that this is the same measurement, but post-processed using two different methods. Both maps show similar magnitudes and trends, with the largest damping in the center and the smallest in the upper right corner. The map of $f_3 - f_2$ also shows a larger linewidth at the lower and right edges. This is, however, because the PLLs need time to settle to the correct frequencies, and the errors are still large there. Although the holes are visible, the picture appears rather smeared out. In contrast, the map made using the estimator $\west$ shows much finer features and does not have artifacts like a too high linewidth at the edges or at the spots where particles are located. Unlike $f_3 - f_2$, $\west$ only needs a short time to settle after the particles. Further detail are shown in Fig.~\ref{fig:supp:wait}. There, the two methods to extract the linewidth are compared for different wait times between points. For short waiting times, $f_3-f_2$ shows a long distance before settling, and no sharp features are seen. By increasing the wait to 10 s per point, the features in the line trace become sharper, but then recording a correct map results in long measurement times. The short waits are also shifted to the right, indicating a delay between the changing linewidth and the reaction of the PLL. In contrast, with $\west$, all traces coincide and show the sharp features, even for the shortest wait of 0.1 s. The estimator thus enables high-resolution mapping and tracking with short measurement times.


\subsection{PLL Asymmetry}
\begin{figure}[tbh]
  \includegraphics[width=.5\columnwidth]{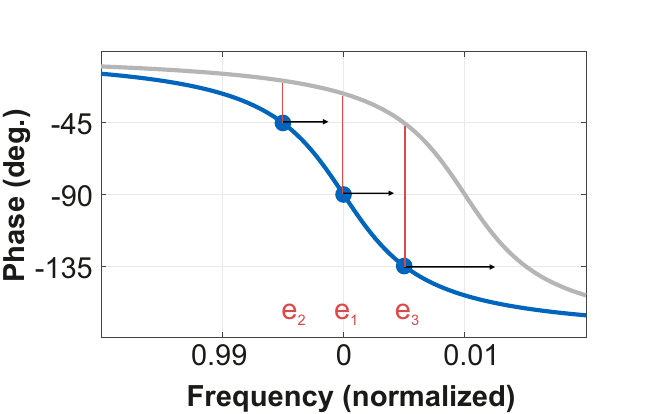}
  \caption{Accumulated errors (indicated in red) of the three PLLs during a frequency shift (gray). The resulting increments by the three  PLLs are indicated in black.
  \label{fig:supp:PID_error}} 
\end{figure}

\noindent The concept behind the PLL tracking is a PI(D) control. Here only the proportional ('P') and integral ('I') terms of the PID are used. For tracking, we lock onto the phase response, 
to a setpoint
, and the PLL aims to maintain a stable phase value, taking past errors into account ('I' term). Suppose that initially the PLL is locked. If the frequency now starts shifting, e.g., due to a change in temperature, the PLL will try to adjust the frequency accordingly. The amount of P-shift is determined from the previous error $e_i$. When tracking three frequencies at the same time, the amount of error is different for each of the PLLs due to the shape of the phase response, see Fig.~\ref{fig:supp:PID_error}, indicated in red. Taking the center-PLL ($-90\degree$) as a reference, the second one at $-45\degree$ has less error in the same amount of time, whereas the third one at $-135\degree$ has more than in the center, leading to different frequency adjustments for each setpoint by  the 'P' term. When the frequency shifts continuously, this situation persists and the 'I' term also contributes. Also this one is smallest for the smaller accumulating errors of the second ($-45\degree$) PLL, at least until the PLLs almost "catch up" and the errors decrease. Thus, an asymmetry arises between the second and third frequency, as was observed in Fig. 3(a).

\section{Operation in the nonlinear regime} \label{supp:nonlinear}
\noindent Looking at Fig.~\ref{fig:supp:FB_setup} shows that the driven response is given by:
\begin{equation}
    y(f) \equiv \frac{V_\mathrm{det}(f)}{V_\mathrm{drive}(f)} = A_\mathrm{piezo} H(f) \tder{V_\mathrm{det}}{u}.
    \label{eq:response}
\end{equation}
This holds for both the linear and nonlinear regime; the difference lies in the resonator response $H(f)$, which is either a Lorentzian (cf. Eq.~\eqref{eq:supp:L}) or is a Duffing response.
We use the following fit function for the driven response $y(f)$\cite{sommer_PRL_nonlinear_map}:
\begin{equation}
    y(f) = \frac{y_\mathrm{max}w/2}{f_0(1 + \frac{3}{8}\alpha_y |y|^2) - f + iw/2} e^{i\psi-2\pi i (f-f_0) \tau}.
    \label{eq:Duffingy}
\end{equation}
The fit parameters are $f_0$, $w$, $y_\mathrm{max}$, $\alpha_y$, $\psi$ and $\tau$. Note the presence of $|y|^2$ in the denominator and that for $\alpha_y =0$, Eq.~\eqref{eq:Duffingy} reduces to a Lorentzian. The parameters $f_0$, $w$, $y_\mathrm{max}$ are the resonance frequency, linewidth and maximum response in the linear regime, respectively. $\psi$ takes the any additional phase at $f_0$ into account and $\tau$ a (group) delay. $\alpha_y$ quantifies the nonlinearity and has units of $\un{(V/V)^{-2}}$. Its critical value $\alpha_c$ is the lowest value of $|\alpha_y|$ where $|y|$ just becomes vertical \cite{cleland_nanomechanics}; beyond that the response becomes hysteretic. In the Lorentzian approximation, its value is $ \alpha_c = \frac{32}{27} \sqrt{3} w / (f_0 y_\mathrm{max}^2)$ \cite{nayfeh_nonlinear}. Hence, once the fit parameters $w$, $f_0$ and $y_\mathrm{max}$ are determined, the value of $\alpha_c$ is fixed. Note that $\alpha_c$ is independent of the driving strength. Still, $\alpha_y$ depends linearly on the driving power $P_\mathrm{drive} \propto V_\mathrm{drive}^2$. Note that in our experiments, there is a stiffening nonlinearity (i.e., $\alpha_y \ge 0$), so this can be written as:
$\alpha_y = \alpha_c P_\mathrm{drive}/P_c = \alpha_c (V_\mathrm{drive}/V_c)^2$, where the latter expression uses the drive voltage amplitude $V_\mathrm{drive}(f)$.
\begin{figure}[tbh]
  \includegraphics[width=.45\textwidth]{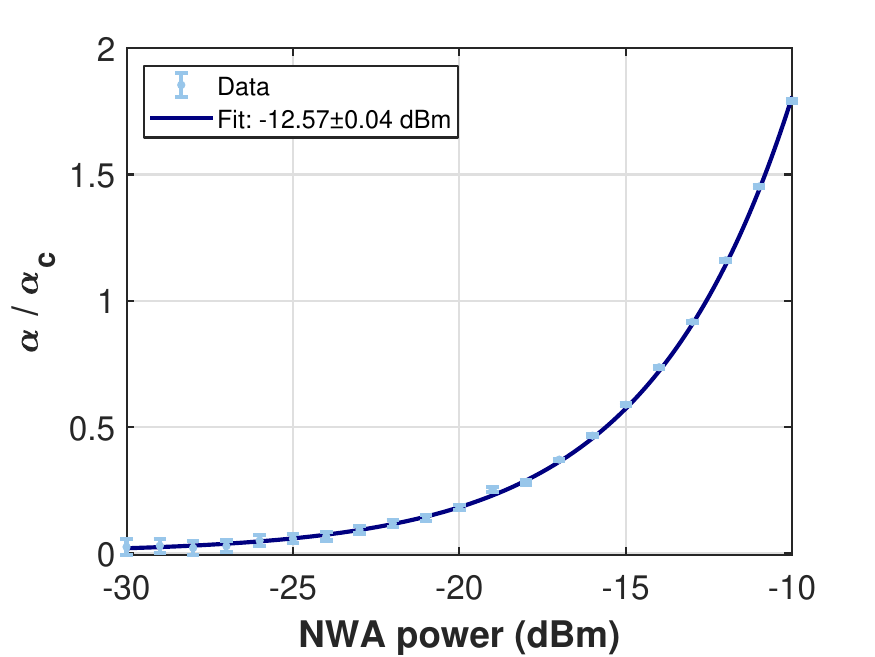}
  \includegraphics[width=.45\textwidth]{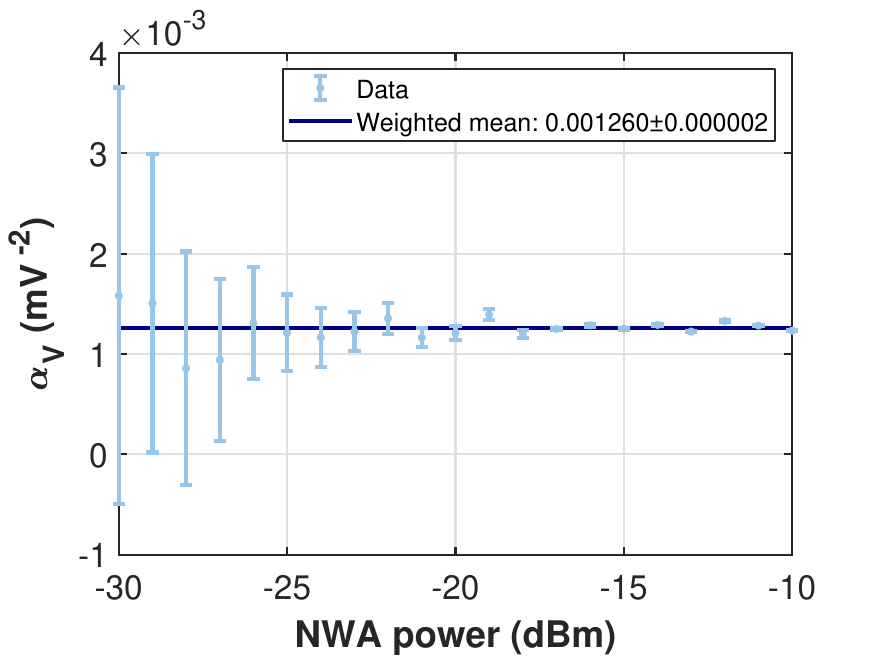} 
  \caption{Normalized Duffing parameter (left) as determined by fitting the phase of the driven responses \cite{sommer_PRL_nonlinear_map} vs. driving power $P_\mathrm{drive}$. A subset of these responses was shown in Fig.~5(a). The solid line is the fitted expected linear (quadratic) dependence on the pump power (amplitude). The right panel shows $\alpha_V$ and its fit uncertainty, together with the weighted mean value.
  \label{fig:supp:alphan}} 
\end{figure}
Figure \ref{fig:supp:alphan} shows the behavior of the fitted normalized Duffing parameter for increasing driving power. The power is varied from $P_\mathrm{drive} = -30\un{dBm}$ to $-10\un{dBm}$ and at every integer value of the drive power, a NWA trace is recorded with the lockin. As the response at high power is influenced by a nonlinear readout, the phase $\angle y(f)$ is used for fitting followed by determining $y_\mathrm{max}$ from the linear part as detailed in Ref.~\onlinecite{sommer_PRL_nonlinear_map}. From the dependence of the normalized Duffing parameter on the driving strength, the critical driving power $P_c$ --- where the normalized Duffing parameter equals 1, i.e. $\alpha_y(P_c ) = \alpha_c$ --- is found at $P_c = -12.57\un{dBm}$. 

For the discussion of the pump-probe experiments in Fig.~5 of the main text, Eq.~\eqref{eq:Duffingy} is rewritten from normalized responses $y(f)$ --- as measured by e.g. an NWA --- to actual voltages: $V_\mathrm{det,pump}(f) = y(f) V_\mathrm{drive,pump}(f)$. For simplicity, we will assume that $\psi = \tau = 0$. Also, it is assumed that the readout is not nonlinear. Then, by expanding the underlying equations of motion for a weak probe tone at frequency $\nu$ around the steady-state solution for the strong pump, the probe response is found \cite{defoort_APL_selfcoupling, Antoni_EPL_nonlinear_membranes}. In the following we use uppercase $V$s for the pump and lowercase $v$s for the weak probe. Their frequencies are labeled $f$ and $\nu$, respectively. Combining all of this, gives:
\begin{eqnarray}
    V_\mathrm{det,pump}(f) & = & \frac{V_\mathrm{max}w/2}{f_0(1 + \hspace{2mm} \frac{3}{8}\alpha_V |V_\mathrm{det,pump}|^2) - f + iw/2} 
    \label{eq:Duffing_pump} \\
    v_\mathrm{det,probe}(\nu) & = & \frac{v_\mathrm{max}w/2}{f_0(1 + 2\frac{3}{8}\alpha_V |V_\mathrm{det,pump}|^2) - \nu + iw/2-\frac{(\frac{3}{8}\alpha_V |V_\mathrm{det,pump}|^2)^2}{f_0(1 + 2\frac{3}{8}\alpha_V |V_\mathrm{det,pump}|^2) - (2f-\nu) - iw/2}}. 
    \label{eq:Duffing_probe}    
\end{eqnarray}
Here, $V_\mathrm{max} = y_\mathrm{max} V_\mathrm{drive,pump}$ and $v_\mathrm{max} = y_\mathrm{max} v_\mathrm{drive,probe}$. Similarly, $\alpha_V = \alpha_y |V_\mathrm{drive,pump}|^{-2}$ is the Duffing parameter in per-Volt-squared, which is independent of the pump strength as $\alpha_y \propto V_\mathrm{drive}^{2}$ as discussed above. This is verified in Fig.~\ref{fig:supp:alphan}.

The equation for the pump [Eq.~\eqref{eq:Duffing_pump}] is just a rewritten version of Eq.~\eqref{eq:Duffingy} and it shows that the detector signal at the pump frequency solely depends on the pump, and not on the weak probe. Now, when using a PLL (``1'') to lock the pump to $\phi_1 = -90\degree$ as is done in the experiments, its frequency ($f \rightarrow f_1$) will track the ``backbone curve'' \cite{Hippold2024} $f_0(1 + \frac{3}{8}\alpha_V |V_\mathrm{max}|^2)$ when changing the pump strength (i.e., $V_\mathrm{drive,pump} \propto V_\mathrm{max}$). This is the straight blue line in Fig.~5(c).

The expression for the probe signal [Eq.~\eqref{eq:Duffing_probe}] is more complicated than that for the pump, but it is noted that the denominator does not contain probe voltages $v$, only pump amplitudes $V$. The former only appears (as $v_\mathrm{max} \propto v_\mathrm{drive,probe}$) in the numerator and $v_\mathrm{det,probe}(\nu)$ is linear in $v_\mathrm{drive,probe}$ \cite{defoort_APL_selfcoupling}.  This linearity also indicates that multiple weak probes (i.e., PLLs 2 and 3) will not influence each other. We define the probe response (in V/V) as $p(\nu) \equiv v_\mathrm{det,probe}(\nu) / v_\mathrm{drive,probe}(\nu)$. Such measured probe responses were shown in Fig.~5(b), whereas a colormap of the phase $\angle p(\nu)$ calculated with Eq.~\eqref{eq:Duffing_probe} vs probe frequency and the normalized Duffing parameter was shown in Fig.~5(c). The PLLs of the weak probes would follow the contour lines (black) corresponding the their setpoints. It is noted that setpoints more negative than the pump are shifted to the right, but never cross, whereas PLLs with less negative setpoints may have to cross the pump frequency. Experimentally, this will give issues with the PLL operation. Likewise, when in- or decreasing the pump strength, a nonmonotonous contour line would require the PLL to make sudden jumps, which is undesirable.

Finally, the peak shape of the probe shows a complicated resonance due to the nature of Eq.~\eqref{eq:Duffing_probe}. The first part of its denominator indicates that the probe probes a Lorentzian resonance at $\nu = f_0(1 + 2\frac{3}{8}\alpha_V |V_\mathrm{det,pump}|^2)$. Its shift is thus twice that of the pump \cite{Antoni_EPL_nonlinear_membranes} and this resonance is intimately related to the libration eigenvalues \cite{houri_PRL_generic_chaos}. In principle, it could be tracked by a probe locked to $-90\degree$. The second part of the denominator is a fraction that is caused by four-wave-mixing processes \cite{defoort_APL_selfcoupling} and has a resonance when $2f -\nu$ matches $f_0(1 + 2\frac{3}{8}\alpha_V |V_\mathrm{det,pump}|^2)$. This term is responsible for the light blue region where $\angle p > 0$ in Fig.~5(c). For a pump locked to $-90\degree$, this term is largest at $\nu = f_0$.

\ifx\calledSI\undefined
    \bibliography{membranes}

@Book{Horowitz,
  author    = {Horowitz, Paul and Hill, Winfield},
  publisher = {Cambridge University Press},
  title     = {The art of electronics},
  year      = {2015},
  edition   = {3rd},
  isbn      = {9780521809269},
  pages     = {1224},
}

@ARTICLE{poot_physrep_quantum_regime,
  author = {Menno Poot and Herre S.J. van der Zant},
  title = {Mechanical systems in the quantum regime},
  journal = {Phys. Rep.},
  year = {2012},
  volume = {511},
  pages = {273--335},
  number = {5},
  doi = {10.1016/j.physrep.2011.12.004},
  issn = {0370-1573},
  url = {http://www.sciencedirect.com/science/article/pii/S0370157311003644}
}

@Article{poot_PRA_squeezing_feedback,
  author    = {Poot, M. and Fong, K. Y. and Tang, H. X.},
  journal   = {Phys. Rev. A},
  title     = {Classical non-Gaussian state preparation through squeezing in an optoelectromechanical resonator},
  year      = {2014},
  month     = dec,
  pages     = {063809},
  volume    = {90},
  doi       = {10.1103/PhysRevA.90.063809},
  issue     = {6},
  numpages  = {6},
  publisher = {American Physical Society},
}

@Article{poot_NJP_Yfeedback,
  author    = {M Poot and K Y Fong and H X Tang},
  journal   = {New J Phys},
  title     = {Deep feedback-stabilized parametric squeezing in an opto-electromechanical system},
  year      = {2015},
  number    = {4},
  pages     = {043056},
  volume    = {17},
  doi       = {10.1088/1367-2630/17/4/043056},
  url       = {http://stacks.iop.org/1367-2630/17/i=4/a=043056},
}

@Article{bleszynski-jayich_science_persistent_currents,
  author    = {Bleszynski-Jayich, A. C. and Shanks, W. E. and Peaudecerf, B. and Ginossar, E. and von Oppen, F. and Glazman, L. and Harris, J. G. E.},
  journal   = {Science},
  title     = {Persistent Currents in Normal Metal Rings},
  year      = {2009},
  month     = oct,
  number    = {5950},
  pages     = {272--275},
  volume    = {326},
  comment   = {10.1126/science.1178139},
  url       = {http://www.sciencemag.org/cgi/content/abstract/326/5950/272},
}

@Article{davidovikj_NL_graphene_mode_visualization,
  author    = {Davidovikj, Dejan and Slim, Jesse J. and Cartamil-Bueno, Santiago J. and van der Zant, Herre S. J. and Steeneken, Peter G. and Venstra, Warner J.},
  journal   = {Nano Lett.},
  title     = {Visualizing the Motion of Graphene Nanodrums},
  year      = {2016},
  issn      = {1530-6984},
  month     = apr,
  number    = {4},
  pages     = {2768--2773},
  volume    = {16},
  comment   = {doi: 10.1021/acs.nanolett.6b00477},
  doi       = {10.1021/acs.nanolett.6b00477},
  publisher = {American Chemical Society},
  url       = {https://doi.org/10.1021/acs.nanolett.6b00477},
}

@Article{poot_APL_cooling,
  author    = {M. Poot and S. Etaki and H. Yamaguchi and H. S. J. van der Zant},
  journal   = {Appl Phys Lett},
  title     = {Discrete-time quadrature feedback cooling of a radio-frequency mechanical resonator},
  year      = {2011},
  month     = {jul},
  number    = {1},
  pages     = {013113},
  volume    = {99},
  doi       = {10.1063/1.3608148},
  eid       = {013113},
  numpages  = {3},
  publisher = {AIP},
}

@Article{sommer_ST_xtalk,
  author  = {Sommer, Timo and Hoch, David and Haas, Kevin-Jeremy and Moller, Leopold and Röwe, Julius and Yadav, Aditya and Soubelet, Pedro and Finley, Jonathan J. and Poot, Menno},
  title   = {Optomechanical mode-shape mapping in the presence of crosstalk},
  journal = {Sensors \& Transducers},
  year    = {2022},
  volume  = {256},
  number  = {4},
  pages   = {1--9},
  month   = jul,
  url     = {http://www.sensorsportal.com/HTML/DIGEST/P_3268.htm},
}

@Article{adiga_APL_SiN_drum_Q_mode,
  author           = {Adiga,V. P. and Ilic,B. and Barton,R. A. and Wilson-Rae,I. and Craighead,H. G. and Parpia,J. M.},
  journal          = {Applied Physics Letters},
  title            = {Modal dependence of dissipation in silicon nitride drum resonators},
  year             = {2011},
  number           = {25},
  pages            = {253103},
  volume           = {99},
  doi              = {10.1063/1.3671150},
  modificationdate = {2023-09-01T16:37:04},
}

@Article{hoch_MM_mode_mapping,
  author         = {Hoch, David and Haas, Kevin-Jeremy and Moller, Leopold and Sommer, Timo and Soubelet, Pedro and Finley, Jonathan J. and Poot, Menno},
  journal        = {Micromachines},
  title          = {Efficient Optomechanical Mode-Shape Mapping of Micromechanical Devices},
  year           = {2021},
  issn           = {2072-666X},
  month          = {jul},
  number         = {8},
  pages          = {880},
  volume         = {12},
  article-number = {880},
  doi            = {10.3390/mi12080880},
  publisher      = {{MDPI} {AG}},
  url            = {https://www.mdpi.com/2072-666X/12/8/880},
}

@Article{Joeckel_PRL_spectroscopy_Q_SiN_membrane,
  author   = {Jöckel, Andreas and Rakher, Matthew T. and Korppi, Maria and Camerer, Stephan and Hunger, David and Mader, Matthias and Treutlein, Philipp},
  journal  = {Appl. Phys. Lett.},
  title    = {Spectroscopy of mechanical dissipation in micro-mechanical membranes},
  year     = {2011},
  issn     = {0003-6951},
  month    = oct,
  number   = {14},
  pages    = {143109},
  volume   = {99},
  doi      = {10.1063/1.3646914},
  url      = {https://doi.org/10.1063/1.3646914},
}

@Article{poot_ST_AlN_simulations,
  author   = {Menno Poot},
  journal  = {Sensors \& Transducers},
  title    = {Sensing the Mechanical Properties of {AlN} Thin Films Using Micromechanical Membranes in Combination with Finite-element Simulations},
  year     = {2023},
  issn     = {2306-8515},
  month    = dec,
  number   = {4},
  pages    = {1-11},
  volume   = {263},
  url      = {https://sensorsportal.com/p_3307.html},
}

@Book{nayfeh_nonlinear,
  author    = {Nayfeh, Ali H.; Mook, Dean T.},
  publisher = {Wiley-VCH},
  title = {Nonlinear Oscillations},
  year = {1979},
  edition   = {1},
  isbn      = {9780471121428},
  month     = mar,
}

@Book{cleland_nanomechanics,
  author           = {A. Cleland},
  publisher        = {Springer},
  title            = {Foundations of Nanomechanics},
  year             = {2003},
  creationdate     = {2024-10-11T13:47:00},
  modificationdate = {2024-10-11T13:47:00},
}

@Article{Antoni_EPL_nonlinear_membranes,
  author           = {Thomas Antoni and Kevin Makles and Rémy Braive and Tristan Briant and Pierre-François Cohadon and Isabelle Sagnes and Isabelle Robert-Philip and Antoine Heidmann},
  journal          = {Europhysics Letters},
  title            = {Nonlinear mechanics with suspended nanomembranes},
  year             = {2013},
  month            = {jan},
  number           = {6},
  pages            = {68005},
  volume           = {100},
  creationdate     = {2025-01-17T12:47:03},
  doi              = {10.1209/0295-5075/100/68005},
  modificationdate = {2025-01-17T12:47:03},
  publisher        = {EDP Sciences, IOP Publishing and Società Italiana di Fisica},
  url              = {https://dx.doi.org/10.1209/0295-5075/100/68005},
}

@Article{Houri_JJAP_pulse_width_duffing,
  author           = {Samer Houri and Ryuichi Ohta and Motoki Asano and Yaroslav M. Blanter and Hiroshi Yamaguchi},
  journal          = {Japanese Journal of Applied Physics},
  title            = {Pulse-width modulated oscillations in a nonlinear resonator under two-tone driving as a means for MEMS sensor readout},
  year             = {2019},
  month            = {feb},
  number           = {SB},
  pages            = {SBBI05},
  volume           = {58},
  creationdate     = {2025-01-17T12:47:03},
  doi              = {10.7567/1347-4065/aaffb9},
  modificationdate = {2025-01-17T12:47:03},
  publisher        = {IOP Publishing},
  url              = {https://dx.doi.org/10.7567/1347-4065/aaffb9},
}

@Article{houri_PRL_generic_chaos,
  author           = {Houri, Samer and Asano, Motoki and Yamaguchi, Hiroshi and Yoshimura, Natsue and Koike, Yasuharu and Minati, Ludovico},
  journal          = {Phys. Rev. Lett.},
  title            = {Generic Rotating-Frame-Based Approach to Chaos Generation in Nonlinear Micro- and Nanoelectromechanical System Resonators},
  year             = {2020},
  month            = {Oct},
  pages            = {174301},
  volume           = {125},
  creationdate     = {2025-01-17T12:47:03},
  doi              = {10.1103/PhysRevLett.125.174301},
  issue            = {17},
  modificationdate = {2025-10-07T16:33:14},
  numpages         = {6},
  publisher        = {American Physical Society},
  url              = {https://link.aps.org/doi/10.1103/PhysRevLett.125.174301},
}

@Article{sommer_APL_membrane_AlN,
  author           = {Sommer, Timo and Aditya and Gross, Rudolf and Althammer, Matthias and Poot, Menno},
  journal          = {Appl. Phys. Lett.},
  title            = {Determining the mechanical properties of {AlN} films using micromechanical membranes},
  year             = {2025},
  issn             = {0003-6951},
  month            = jan,
  number           = {4},
  pages            = {043508},
  volume           = {126},
  doi              = {10.1063/5.0237655},
  modificationdate = {2025-04-15T12:24:03},
  url              = {https://doi.org/10.1063/5.0237655},
}

@Article{Shaniv_PRR_brownian,
  author           = {Shaniv, Ravid and Reetz, Chris and Regal, Cindy A.},
  journal          = {Physical Review Research},
  title            = {Direct measurement of a spatially varying thermal bath using {Brownian} motion},
  year             = {2023},
  issn             = {2643-1564},
  month            = nov,
  number           = {4},
  pages            = {043121},
  volume           = {5},
  creationdate     = {2025-04-03T13:04:33},
  doi              = {10.1103/physrevresearch.5.043121},
  modificationdate = {2025-11-25T10:12:48},
  publisher        = {American Physical Society (APS)},
}

@Misc{catalini_arXiv_detuned_tones,
  author           = {Letizia Catalini and Javier del Pino and Soumya S. Kumar and Vincent Dumont and Gabriel Margiani and Oded Zilberberg and Alexander Eichler},
  title            = {Slow and fast topological dynamical phase transitions in a Duffing resonator driven by two detuned tones},
  year             = {2024},
  archiveprefix    = {arXiv},
  creationdate     = {2025-04-07T16:36:52},
  eprint           = {2408.15794},
  modificationdate = {2025-04-07T16:37:22},
  primaryclass     = {cond-mat.mes-hall},
  url              = {https://arxiv.org/abs/2408.15794},
}

@Article{sommer_PRL_nonlinear_map,
  author           = {Sommer, Timo and Zinth, Agnes and Aditya and Poot, Menno},
  journal          = {Phys. Rev. Lett.},
  title            = {Spatial Mapping of Intrinsic and Readout Nonlinearities in a Strongly Driven Micromechanical Membrane},
  year             = {2025},
  month            = {May},
  pages            = {203604},
  volume           = {134},
  creationdate     = {2025-05-30T11:45:00},
  doi              = {10.1103/PhysRevLett.134.203604},
  issue            = {20},
  modificationdate = {2025-05-30T11:45:00},
  numpages         = {6},
  publisher        = {American Physical Society},
  url              = {https://link.aps.org/doi/10.1103/PhysRevLett.134.203604},
}

@Article{Hippold2024,
  author    = {Hippold, Patrick and Scheel, Maren and Renson, Ludovic and Krack, Malte},
  journal   = {Mechanical Systems and Signal Processing},
  title     = {Robust and fast backbone tracking via phase-locked loops},
  year      = {2024},
  issn      = {0888-3270},
  month     = nov,
  pages     = {111670},
  volume    = {220},
  doi       = {10.1016/j.ymssp.2024.111670},
  publisher = {Elsevier BV},
}

@PhdThesis{hoch_thesis,
  author           = {David Christopher Julian Hoch},
  school           = {TU Munich},
  title            = {Optomechanics with high-stress silicon nitride resonators},
  year             = {2023},
  creationdate     = {2025-10-22T09:56:24},
  modificationdate = {2025-10-22T09:59:05},
}

@book{cleland2013foundations,
  title={Foundations of nanomechanics: from solid-state theory to device applications},
  author={Cleland, Andrew N},
  year={2013},
  publisher={Springer Science \& Business Media}
}

@article{denis2018identification,
  title={Identification of nonlinear modes using phase-locked-loop experimental continuation and normal form},
  author={Denis, Vivien and Jossic, M and Giraud-Audine, Christophe and Chomette, B and Renault, A and Thomas, Olivier},
  journal={Mechanical Systems and Signal Processing},
  volume={106},
  pages={430--452},
  year={2018},
  publisher={Elsevier}
}

@article{blaikie2019fast,
  title={A fast and sensitive room-temperature graphene nanomechanical bolometer},
  author={Blaikie, Andrew and Miller, David and Alem{\'a}n, Benjam{\'\i}n J},
  journal={Nature communications},
  volume={10},
  number={1},
  pages={4726},
  year={2019},
  publisher={Nature Publishing Group UK London}
}

@Article{zinth_membrane_readout,
  author           = {Agnes Zinth, Menno Poot},
  journal          = {in preparation},
  title            = {Reconstruction and Analysis of Vibrational Mode Shapes in Micromechanical Membranes},
  year             = {2025},
}

@article{sanz2022high,
  title={High-throughput determination of dry mass of single bacterial cells by ultrathin membrane resonators},
  author={Sanz-Jim{\'e}nez, Adri{\'a}n and Malvar, Oscar and Ruz, Jose J and Garc{\'\i}a-L{\'o}pez, Sergio and Kosaka, Priscila M and Gil-Santos, Eduardo and Cano, {\'A}lvaro and Papanastasiou, Dimitris and Kounadis, Diamantis and Mingorance, Jes{\'u}s and others},
  journal={Communications Biology},
  volume={5},
  number={1},
  pages={1227},
  year={2022},
  publisher={Nature Publishing Group UK London}
}

@inproceedings{aubin2003laser,
  title={Laser annealing for high-Q MEMS resonators},
  author={Aubin, Keith L and Zalalutdinov, Maxim and Reichenbach, Robert B and Houston, Brian H and Zehnder, Alan T and Parpia, Jeevak M and Craighead, Harold G},
  booktitle={Smart Sensors, Actuators, and MEMS},
  volume={5116},
  pages={531--535},
  year={2003},
  organization={SPIE}
}

@article{jaeger2023mechanical,
  title={Mechanical mode imaging of a High-Q hybrid hBN/Si3N4 resonator},
  author={Jaeger, David and Fogliano, Francesco and Ruelle, Thibaud and Lafranca, Aris and Braakman, Floris and Poggio, Martino},
  journal={Nano Letters},
  volume={23},
  number={5},
  pages={2016--2022},
  year={2023},
  publisher={ACS Publications}
}

@article{guneroglu2025study,
  title={Study of Frequency Trimming Ability and Performance Enhancement of Thin-Film Piezoelectric-on-Silicon MEMS Resonators by Joule Heating via Localized Annealing},
  author={Guneroglu, Ugur and Zaman, Adnan and Alsolami, Abdulrahman and Rivera, Ivan F and Wang, Jing},
  journal={IEEE Transactions on Ultrasonics, Ferroelectrics, and Frequency Control},
  year={2025},
  publisher={IEEE}
}

@article{liu2022effect,
  title={The effect of annealing and optical radiation treatment on graphene resonators},
  author={Liu, Yujian and Li, Cheng and Fan, Shangchun and Song, Xuefeng and Wan, Zhen},
  journal={Nanomaterials},
  volume={12},
  number={15},
  pages={2725},
  year={2022},
  publisher={MDPI}
}

@article{vsivskins2020magnetic,
  title={Magnetic and electronic phase transitions probed by nanomechanical resonators},
  author={{\v{S}}i{\v{s}}kins, Makars and Lee, Martin and Ma{\~n}as-Valero, Samuel and Coronado, Eugenio and Blanter, Yaroslav M and van der Zant, Herre SJ and Steeneken, Peter G},
  journal={Nature communications},
  volume={11},
  number={1},
  pages={2698},
  year={2020},
  publisher={Nature Publishing Group UK London}
}

@article{sader2018mass,
  title={Mass spectrometry using nanomechanical systems: beyond the point-mass approximation},
  author={Sader, John E and Hanay, M Selim and Neumann, Adam P and Roukes, Michael L},
  journal={Nano letters},
  volume={18},
  number={3},
  pages={1608--1614},
  year={2018},
  publisher={ACS Publications}
}

@article{bhattacharya2025two,
  title={Two Dimensional-Material-Coated Microcantilevers for Enhanced Mass Sensing and Material Characterization},
  author={Bhattacharya, Gourav and Lionadi, Indrianita and Taverne, Mike and McLaughlin, James and Fernandez-Ibanez, Pilar and Huang, Chung-Che and Ho, Ying-Lung Daniel and Payam, Amir Farokh and others},
  journal={Nanoscale},
  year={2025},
  publisher={Royal Society of Chemistry}
}

@article{bozkurt2025mechanical,
  title={A mechanical quantum memory for microwave photons},
  author={Bozkurt, Alk{\i}m B and Golami, Omid and Yu, Yue and Tian, Hao and Mirhosseini, Mohammad},
  journal={Nature Physics},
  pages={1--6},
  year={2025},
  publisher={Nature Publishing Group UK London}
}

@article{kosaka2014detection,
  title={Detection of cancer biomarkers in serum using a hybrid mechanical and optoplasmonic nanosensor},
  author={Kosaka, Priscila M and Pini, Valerio and Ruz, Jose Jaime and Da Silva, RA and Gonz{\'a}lez, MU and Ramos, Daniel and Calleja, Montserrat and Tamayo, Javier},
  journal={Nature nanotechnology},
  volume={9},
  number={12},
  pages={1047--1053},
  year={2014},
  publisher={Nature Publishing Group UK London}
}

@article{arlett2011comparative,
  title={Comparative advantages of mechanical biosensors},
  author={Arlett, JL and Myers, EB and Roukes, ML},
  journal={Nature nanotechnology},
  volume={6},
  number={4},
  pages={203--215},
  year={2011},
  publisher={Nature Publishing Group UK London}
}

@article{li2024miniature,
  title={Miniature optical fiber photoacoustic spectroscopy gas sensor based on a 3D micro-printed planar-spiral spring optomechanical resonator},
  author={Li, Taige and Zhao, Pengcheng and Wang, Peng and Krishnaiah, Kummara Venkata and Jin, Wei and Zhang, A Ping},
  journal={Photoacoustics},
  volume={40},
  pages={100657},
  year={2024},
  publisher={Elsevier}
}

@article{kort2022utilization,
  title={Utilization of coupled eigenmodes in Akiyama atomic force microscopy probes for bimodal multifrequency sensing},
  author={Kort-Kamp, Wilton JM and Murdick, Ryan A and Htoon, Han and Jones, Andrew C},
  journal={Nanotechnology},
  volume={33},
  number={45},
  pages={455501},
  year={2022},
  publisher={IOP Publishing}
}

@article{santos2023quantification,
  title={Quantification of van der Waals forces in bimodal and trimodal AFM},
  author={Santos, Sergio and Gadelrab, Karim and Elsherbiny, Lamiaa and Drexler, Xaver and Olukan, Tuza and Font, Josep and Barcons, Victor and Chiesa, Matteo},
  journal={The Journal of chemical physics},
  volume={158},
  number={20},
  year={2023},
  publisher={AIP Publishing}
}

@article{chandrashekar2022sensitivity,
  title={Sensitivity of viscoelastic characterization in multi-harmonic atomic force microscopy},
  author={Chandrashekar, Abhilash and Givois, Arthur and Belardinelli, Pierpaolo and Penning, Casper L and Arag{\'o}n, Alejandro M and Staufer, Urs and Alijani, Farbod},
  journal={Soft matter},
  volume={18},
  number={46},
  pages={8748--8755},
  year={2022},
  publisher={Royal Society of Chemistry}
}

@Article{defoort_APL_selfcoupling,
  author           = {Defoort, M. and Lulla, K. J. and Blanc, C. and Bourgeois, O. and Collin, E. and Armour, A. D.},
  journal          = {Applied Physics Letters},
  title            = {Modal “self-coupling” as a sensitive probe for nanomechanical detection},
  year             = {2013},
  issn             = {0003-6951},
  month            = {07},
  number           = {1},
  pages            = {013104},
  volume           = {103},
  creationdate     = {2025-11-12T15:03:33},
  doi              = {10.1063/1.4812718},
  eprint           = {https://pubs.aip.org/aip/apl/article-pdf/doi/10.1063/1.4812718/13207883/013104\_1\_online.pdf},
  modificationdate = {2025-11-12T15:03:33},
  publisher        = {AIP Publishing},
  url              = {https://doi.org/10.1063/1.4812718},
}

@Article{yang_NL_masssensing,
  author           = {Yang, Y. T. and Callegari, C. and Feng, X. L. and Ekinci, K. L. and Roukes, M. L.},
  journal          = {Nano Letters},
  title            = {Zeptogram-Scale Nanomechanical Mass Sensing},
  year             = {2006},
  issn             = {1530-6984},
  month            = apr,
  number           = {4},
  pages            = {583--586},
  volume           = {6},
  comment          = {doi: 10.1021/nl052134m},
  creationdate     = {2025-11-25T16:20:33},
  modificationdate = {2025-11-25T16:20:33},
  publisher        = {American Chemical Society},
  url              = {http://dx.doi.org/10.1021/nl052134m},
}

@Article{Garcia-Sanchez_PRL_Casimir,
  author           = {Garcia-Sanchez, Daniel and Fong, King Yan and Bhaskaran, Harish and Lamoreaux, Steve and Tang, Hong X.},
  journal          = {Phys. Rev. Lett.},
  title            = {Casimir Force and In Situ Surface Potential Measurements on Nanomembranes},
  year             = {2012},
  month            = {Jul},
  pages            = {027202},
  volume           = {109},
  creationdate     = {2025-11-25T16:25:14},
  doi              = {10.1103/PhysRevLett.109.027202},
  issue            = {2},
  modificationdate = {2025-11-25T16:25:14},
  numpages         = {5},
  publisher        = {American Physical Society},
  url              = {https://link.aps.org/doi/10.1103/PhysRevLett.109.027202},
}
    \end{document}
\fi

  \fi
  \end{document}
\fi
%